
\magnification\magstep 1
\baselineskip=0.65 true cm

\vsize 24.0true cm

\def\lsim{\mathrel{\rlap{\lower4pt\hbox{\hskip1pt$\sim$}}
    \raise1pt\hbox{$<$}}}         
\def\gsim{\mathrel{\rlap{\lower4pt\hbox{\hskip1pt$\sim$}}
    \raise1pt\hbox{$>$}}}         

\def\build#1_#2^#3{\mathrel{
\mathop{\kern 0pt#1}\limits_{#2}^{#3}}}
\def\x0{\build{\hbox to 10mm{\sim}}_{x\rightarrow 0}^{ }}


\def\NP#1#2#3{ Nucl. Phys. {\bf #1} (#2) #3}
\def\NPps#1#2#3#4{ Nucl. Phys. #1 (Proc. Suppl.) {\bf #2} (#3) #4}

\def\PL#1#2#3{ Phys. Lett. {\bf #1} (#2) #3}

\def\PR#1#2#3{ Phys. Rev. {\bf #1} (#2) #3}

\def\ZP#1#2#3{ Z. Phys. {\bf #1} (#2) #3}

\def\SJNP#1#2#3{ Sov. J. Nucl. Phys.  {\bf #1} (#2) #3}
\def\JETP#1#2#3{ Sov. Phys. JETP {\bf #1} (#2) #3}

\def\NC#1#2#3{ Nuovo Cimento {\bf #1} (#2) #3}


\centerline{\bf}
\vskip 24pt

\centerline{\bf DO WE NEED TWO POMERONS?}
\vskip 3 true cm

\centerline{M. BERTINI, \hskip 0.5 true cm   M. GIFFON}
\smallskip

\centerline{Institut de Physique Nucl\'eaire de Lyon }
\centerline{Universiti Claude Bernard Lyon - France}

\vskip 0.5 true cm

\centerline{and}

\vskip 0.5 true cm

\centerline{E. PREDAZZI}
\smallskip
\centerline{ Dipartimento di Fisica Teorica dell' Universit\`a di Torino -
Italy}
\centerline{ and Sezione INFN di Torino - Italy}

\vskip 3 true cm

\noindent{\bf Summary.} We show that one single Pomeron compatible with
the Froissart limit, can account for all the present HERA data.

\vfill\eject


High energy diffraction, popular some twenty years ago in hadronic physics,
has been rejuvenated after many years of almost total neglect by the so-called
low-$x$ physics {\it i.e.} by the measurement at HERA of
the proton structure function $\nu W_2$ at small
$x$ [1,2]. A terminology which had become nearly obsolete is essentially
being rediscovered and of great interest is presently the connection between
this new physics and the traditional high energy hadronic
physics. The main issue at stake is whether QCD may shed light on the
origin and the nature of the Pomeron, the entity which, in the conventional
language of high energy physics determines the asymptotic behavior of
 the hadronic total cross sections. More specifically, the question is the
precise
determination of the Pomeron structure function following the original
suggestions of Ingelman and Schlein [3] and of Donnachie and Landshoff [4].
It is not our aim in this paper neither to review the (by now fairly large)
literature on this subject [5], nor to debate how
much precisely gluonic or partonic components the data seem to attribute
to the Pomeron according to the various analyses [6] nor
how well the data are accounted for by the various models [7].
Similarly, it is not our goal to review and update the old fashioned
terminology (see for instance Ref. 8). What we want to do in this
paper is to challenge the rather widespread belief that {\it two Pomerons}
are necessary to describe the physical situation (even though the philosophies
in these two papers are profoundly different, the reader could benefit from
reading, for instance, the papers quoted in Ref. 9a, and 9b).

We will try to reduce the formulation of our problem to its bare minimum at
the risk of oversimplifying it (the kinematic and the variables to be used
are perfectly conventional from Deep Inelastic Scattering (DIS) and summarized
in Fig. 1 for the reaction $\ell(k) + N(p) \rightarrow \ell'(k')) + N(p') + X$
where $\ell$ is a lepton, $N$ is a nucleon and $X$ is all the remaining
hadronic
debris over whose variables a summation is implied).

\item {i)} When $Q^2 \rightarrow 0$, $\nu W_2$ is related to the total
cross section for real-photon proton scattering according to :

$${\sigma}_{tot}^{\gamma p} \, = \, {{4 \pi{^2} \alpha} \over {Q^2}} \nu
W_2|_{Q^2=0}\ ,\eqno(1)$$

(where $\alpha$ is the electromagnetic coupling constant) as a consequence
$\nu W_2$ must vanish

linearly with $Q^2$.

\item {ii)} When the Bjorken variable of DIS $x$ is {\it very} small (say,
typically $x \leq 10^{-3}$), $\nu W_2$ will be dominated by its gluonic
component
and we are going to assume this even when comparing our form
with data at considerably larger $x$, say of order $10^{-2}$.
In this kinematical range we will run perilously close to where our
approximation may break down; on the other hand the complications we would
have to introduce to avoid this danger would make our analysis much more
muddy and, consequently, much less conclusive.

\item {iii)} According to the conventional Regge theory, the asymptotic
behavior of hadronic cross sections as $s \rightarrow \infty$
should be {\it up to logarithms} of the form :

$${\sigma}_{tot}
\build{\hbox to 8mm{\rightarrowfill}}
_{s\rightarrow\infty} ^{ }
s^{(\alpha(0)-1)}\ ,\eqno(2)$$

\item{}
where $\alpha(0)$ is the {\it intercept} of each contributing Regge
trajectory (of which, when the quantum numbers are those of the
vacuum, the dominant, $\alpha_P(0)$ is known as the {\it Pomeron intercept}).

\item {iv)} For a diffractive process (such as the one analyzed at HERA,
$e + p \rightarrow e' + p' + X$ where X has the quantum numbers of a vector
meson), the dominant contribution comes from the Pomeron for which the
intercept is allowed to attain its maximum value compatible with unitarity
$\alpha_P(0)=1.$
In this case, however, logarithmic contributions are expected in Eq.(2)
but, let us stress,

\item {v)} unitarity guarantees that it must be :
$$\alpha_P(0) \leq 1\ .\eqno(3)$$

In particular, Froissart's bound [10] states that a hadronic total
cross section cannot grow faster than $\ln^2 s$. Translated into the language
of structure functions, owing to the correspondence :

$$W^2 \, = \, M^2 \, + \, {Q^2} {{1-x} \over {x}} \ ,\eqno(4)$$

\noindent
(where $W^2$, the total squared energy of the system ${\gamma}^* p$ is the
equivalent of $s$ in a hadronic reaction), Froissart bound states that,
asymptotically, as $W^2 \rightarrow \infty$,
{\it i.e.} as $1/x \rightarrow \infty$ :

$$({\nu W_2 \over{Q^2}})
\propto \ln^2(1/x)\ . \eqno(5)$$


\noindent In what follows, we will show that one can indeed accomodate the HERA
data to this limiting logarithmic behavior (or to a
$\ln (1/x)$ one), in the line of thought of Ref. [11], instead of the power one
discussed below (eq.(8)).

Concerning this latter point, it was, in fact, shown long ago by
Donnachie and Landshoff [12] that an {\it effective} Pomeron intercept of :

$$\alpha_P(0)=1.08\ ,\eqno(6)$$
{\it i.e.} an {\it effective} form of the total cross sections :

$$\sigma_{tot}(s \to \infty) \propto s^{\epsilon} \quad \quad
{\hbox{{\tenrm where}}} \quad  \epsilon \approx 0.08\ ,\eqno(7)$$

\noindent
accounts very well for a large quantity of data. Eq. (7) formally
violates Froissart bound but the idea is that this will occur only at
fantastically high energies\footnote{$^*$}{ One should worry, however, not only
about the violation of Froissart's bound but also of the S-wave unitarity.}
which will probably never be reached and where, presumably, higher order
corrections (such as multi-Pomeron cuts) will restore the validity of
Froissart's
bound. Be as it may, the point is that the form (7) is phenomenologically
quite adequate and with a minimum of parameters accounts, qualitatively, for
a large set of data. Moreover, as shown by the same authors, the combination
(1+7) extrapolates well the photoproduction cross section to the HERA energy
domain. More precisely, one can say that it
accounts well for the early HERA data (in Figure 2 which is taken from Ref. 9a,
these data, not shown would lie along the curve up to $x$ not smaller
than some $10^{-2}$). Actually, the form which is shown in Figure 2
corresponds to including subasymptotic corrections suggested by the
Regge pole analysis, {\it i.e.} the curve is :

$$\nu W_2 \, = \, 0.32 \, x^{-0.08} \, ({{Q^2} \over {Q^2+a}})^{1.08} \, + \,
0.10 \, x^{0.45} \, ({{Q^2} \over {Q^2+b}})^{0.55} \  , \eqno(8)$$
where
$$a \, = \, (750\
{\hbox{{\tenrm MeV}}})^2 \quad \quad \quad \quad
b \, = \, (110\ {\hbox{{\tenrm MeV}}})^2.$$

As one sees from Fig. 2, however, Eq. (8) while reproducing well the data for
$x$ not smaller than $\approx 10^{-2}$ and $Q^2$ small, fails quite badly
when extrapolated to much smaller $x$ values where the latest HERA data
show a much sharper rise.

Two problems arise at this point. One, conceptual, is, could this treatment
be extended to the case in which the Froissart bound is respected ( {\it i.e.}
could we use a form which would behave as (5) in the proper domain)? and the
second, practical one, is, can this treatment be made compatible with the
ensemble of HERA data {\it
small $x$ but large $Q^2$} which, on the contrary, deviate
drastically from the form (8)?

These questions are central to our present paper. Concerning the second,
practical point, this is precisely the reason why, in the literature, one
introduces  [9,13] something which we will call a
{\it hard Pomeron} [14] in order to recover agreement with the data.
On the other hand, always concerning this point, doubts about the real
necessity of doing so are raised by some recent findings [15].

It is our contention that the conclusion that {\it two Pomerons},
a {\it hard} Pomeron\footnote{$^+$}{The intercept of a hard Pomeron would be
somewhere between 0.3 and 0.5 {\it i.e.} much larger than the value 0.08 of
Eq. (6). This is why the case of Eq. (8) is also referred to as a {\it soft
Pomeron} in the literature.}, and a {\it soft} Pomeron,
to simplify somehow the issue are necessary, is
not really required by the data and that one can live without this somewhat
disturbing if not directly unpleasant possibility.

A very interesting way out was suggested recently by Capella et al.[16],
that  the Pomeron intercept  could have
a $Q^2$ dependence. In Ref. 16, however, this possibility was exploited to
obtain a {\it soft Pomeron i.e.} \`a la Donnachie-Landshoff
starting from a {\it hard
Pomeron} \`a la Lipatov {\it et al.} In this paper, rather than
using two components for the Pomeron
(describing its  small $Q^2$ and large
$Q^2$ contributions to the structure functions as in Ref. 11b)
we wish to suggest that both points, the conceptual violation of unitarity
by the {\it soft Pomeron} of Eq. (6) and the practical one, {\it i.e.} a good
reproduction of HERA data, could be offered by an extension of the method
suggested in Ref. 16 by allowing the Pomeron intercept to vary with $Q^2$
in such a way that in the limit $Q^2 \rightarrow 0$ the {\it Froissart Pomeron
i.e.} a $\ln^2(1/x)$ form (see Eq. (5)) is obtained.

To make our point, we propose a specific {\it small x} form for $\nu W_2$ which
{\it i)} fits well all the {\it small x} HERA data and {\it ii)} reduces to
a form (5) (or, alternatively to a $\ln(1/x)$) limit when
$Q^2 \rightarrow 0$.
Specifically, we propose, as an example (certainly other examples could be
offered) :

$$\nu W_2(x,Q^2)
\simeq
A_P \bigg[
{{\tilde x}^{\epsilon(Q^2)}-(1+\epsilon(Q^2)\ln(\tilde x))
\over{{1\over{2}}\epsilon^2(Q^2)}}\bigg]
\ln \bigg(1+{Q^2\over{Q^2+{a}^2_{Pom}}}\bigg)
\ ,\eqno(9)$$

\noindent
or, alternatively :

$$ \nu W_2(x,Q^2)
\simeq
A_P \bigg[
{{\tilde x}^{\epsilon(Q^2)}-1 \over{{\epsilon(Q^2)}}}\bigg]
\ln \bigg(1+{Q^2\over{Q^2+{a}^2_{Pom}}}\bigg)
\ , \eqno(10)$$

\noindent where $\tilde x=W^2/s_0$, with the hadronic scale taken as
$s_0=1$ GeV$^2$.

These forms reduce to the wanted cases if $\epsilon (Q^2)$ vanishes as
$Q^2 \rightarrow 0$ because $\ln(\tilde x)\simeq \ln(1/x)$ if $W^2 \gg Q^2$
. Again as an example, in both cases, we choose for the
intercept $\epsilon (Q^2)$ the specific (and arbitrary) form :

$$ \epsilon(Q^2)={\lambda\over{\ln 2}}
\ln \bigg(1+{Q^2\over{Q^2+{b}^2}}\bigg)\ , \eqno(11)$$
which we borrow from Ref.[11b].
\noindent Then Eq. (9) leads to a $\ln^{2}(1/x)$ behavior and Eq. (10) to a
$\ln(1/x)$.

In Eqs. (10,11) the parameters $A_P$ and ${a}^2_{Pom}$ are fixed by the
requirement that the total photoproduction cross section comes out correct. We
take the values obtained from previous results on  ${\sigma}^{\gamma p}_{tot}$
[11b].

So, with the specific choice (11) of $\epsilon (Q^2)$ there are just
two adjustable parameters $\lambda$ and $b^2$. Fitting the
{\it small x} (specifically up to $x \leq 5 \times 10^{-3}$), the result is
shown for the case of Eq. (9) in Fig.3 and the best fit to the parameters
gives $A_P=5.72\ 10^{-3}$, ${a}^2_{Pom}=1.12 $ GeV$^2$, $\lambda=0.254$, and
$b^2=0.198$ GeV$^2$
with a $\chi ^2$(/d.o.f)(/58 HERA data) of about 1.2.

The result of Fig. 3 is quite spectacular and deserves some comments (the NMC
data [17], not fitted, are shown for completeness). First,
recall that the data with $x \geq 5 \times 10^{-3}$ are {\it not} the result
of a best fit; in spite of this, it is only for very high $Q^2$ that the curve
deviates considerably from the data. Second, had we used Eq. (10) instead of
Eq. (9),
the result would have been quite similar. Third and perhaps most
interesting, notice that the asymptotic value of $\epsilon$ as $Q^2$ grows to
$ \approx 2000$ GeV$^2$ is, roughly $=0.3$ {\it i.e.} reaches the lower limit
of what are considered the range of values appropriate for the {\it hard
Pomeron} (the value of the {\it soft Pomeron} \`a la Donnachie and Landshoff,
0.08, being reached for $Q^2$ between 1 and 5  GeV$^2$). Notice also, that no
evolution \`a la Altarelli-Parisi has been taken into account to get
the previous results (to perform a correct evolution, the whole machinery of
structure functions, of their gluonic and of their partonic contributions
would have to be properly taken into account. This, however, would obvioulsy
improve the fit but would make the result depend on so many additional facts
and parameters that the main point
of the paper would be lost in the details of the parametrization).

In order to see what happens when a factor correcting for $x$ not being so
small
is inserted into Eq. (9) (or (10)), we show in Fig. 4 the result obtained
repeating the previous procedure with the form :

$$
\nu W_2(x,Q^2)\simeq
A_P \bigg[
{{\tilde x}^{\epsilon(Q^2)}-(1+\epsilon(Q^2)\ln(\tilde x))
\over{{1\over{2}}\epsilon^2(Q^2)}}\bigg]
\ln \bigg(1+{Q^2\over{Q^2+{a}^2_{Pom}}}\bigg)
(1-x)^{\beta(Q^2)}
\ ,\eqno(12)$$

\noindent where,

$$
\beta(Q^2)=\beta_0+\beta_1 t
\qquad {\hbox{{\tenrm{with}}}\qquad
t=\ln \bigg(
{\ln \big((Q^2+{Q}_0^2)/{\Lambda}^2 \big)
\over{\ln \big({Q}_0^2/{\Lambda}^2 \big)}}} \bigg) \ ,
\eqno(13)$$

\noindent
(the same form (11) has beeen used for $\epsilon (Q^2)$).
Fig. 4a (obtained with the form (12)) shows the
equivalent of Fig. 3 {\it i.e.} the structure function as a function of $ x$
for the various available bins in $Q^2$ whereas Fig. 4b shows the converse
{\it i.e.} the variation in $Q^2$ for the various bins in $x$. Compared with
the previous result,
the $\chi ^2 $ (/d.o.f) (/67  HERA data )
is now 1.55 and the various parameters are
now given by:
$A_P=5.72\ 10^{-3}$, ${a}^2_{Pom}=1.12 $ GeV$^2$, $\lambda=0.256$, and
$b^2=0.21$ GeV$^2$, $\beta_0=7.0$, $\beta_1=5.6$.
As expected, the overall picture has further improved
proving that the large $x$ disagreement in Fig. 3 was largely due to the lack
of an appropriate treatment of the not so small $x$ data (in Fig. 4  HERA
data  for $x \le 10^{-2}$ have been fitted, not just those
below $x \approx 5 \times 10^{-3}$ as in the previous case).
Notice also that the parameters already present in the previous fit have
practically remained the same since they
were determined to reproduce the small-$x$ data;
only the parameters involved in $\beta(Q^2)$ are sensitive to including
larger $x$-values in the fit.

Once again, $\epsilon (Q^2)$ is closed to $0.3$ at the highest $Q^2$ values
and crosses the {\it soft} value 0.08 for $Q^2$ somewhere between 1 and 5
GeV$^2$.

Some general conclusions are in order. We have shown that we can live well
{\it without two Pomerons} and, furthermore, that a form compatible with
the Froissart limit, which we call {\it Froissart Pomeron} is quite
acceptable. The form we offered is quite {\it ad hoc} but this is true of
basically all the parametrizations used in this game. No doubt more clever
and elaborate forms could be offered and, no doubt, the analysis could be
largely ameliorated, for example by using the whole machinery in which not
only gluon distributions are taken into account but also partons together,
of course, with their correct $Q^2$ evolution. This, however, raises the
issue of how well one could fit the ensemble of all data on structure
functions with a parametrization of the kind proposed here.
We hope to come back to these questions in the future.

\vfill \eject

\centerline{\bf References}
\medskip

\item{[1]} H1 Collaboration, I. Abt {\it et al.}, Nucl. Phys. {\bf B407}
(1993) 515.
\smallskip

\item{[2]} ZEUS Collaboration, M. Derrick {\it et al.}, Phys. Lett.
{\bf B316} (1993) 412.

\item{} ZEUS Collaboration  {\it Measurement of the Proton Structure Function
$F_2$ from the 1993 HERA Data} DESY 94-113 August 1994.
\smallskip

\item {[3]} G. Ingelman and P. Schlein, Phys. Lett. {\bf B152} (1985) 256.
\smallskip

\item {[4]} A. Donnachie and P. V. Landshoff, Phys. Lett. {\bf B191} (1987)
309 and Nucl. Phys. {\bf B303} (1988) 634.

\item {[5]} See, for instance,
P.V Landshoff,  XXXVII Rencontres de Moriond, March 1992 Ed. by J.
Tran Than Van (Edition Frontihres) and references therein; see also

\item{}G. Ingelman and K. Prytz, \ZP{C58}{1993}{285};

\item{}G. Ingelman, J. Phys G {\bf 19} {\it Workshop on HERA - the New Frontier
for QCD} (1994) 1631.
\smallskip

\item {[6]} UA1 Collaboration K. Eggert, $2^{th}$ Blois workshop on Elastic and
Diffractive Scattering, Rockefeller University, New York, USA, 1987;

\item{} UA8 Collaboration, P.E Schlein,  \NPps{33A, B}{1993}{41};

\PL{B332}{1994}{126}.
\smallskip

\item {[7]} P.V Landshoff and O. Nachtman,  \ZP{C35}{1987}{405};

\item{}D.A. Ross,  J.Phys G {\bf 15} (1989) 1175;

\item{}N.N Nikolaev and B.G Zakharov,  \ZP{C53}{1992}{331};

\item{} E. Gotsman, E.M Levin and U. Maor,  \ZP{C57}{1993}{677};

\item{}J.R Cudell and B.U Nguyen, \NP{B420}{1994}{669}.
\smallskip

\item {[8]} E. Predazzi,  {\it Perspectives in High Energy Physics} Lectures
delivered at the

III$^{th}$ G. Wataghin School in Phenomenology, Campinas, July
1994.
\smallskip

\item {[9]} a) P. V. Landshoff, {\it The Two Pomerons} Lecture delivered at the
PSI school at Zuoz,

August 1994;

\item{} A. Donnachie and P.V. Landshoff, \ZP{C61}{1994}{139}.
\item{} b) M. Genovese, N. N. Nikolaev and B. G. Zakharov, {\it Diffractive
DIS from the Generalized BFKL Pomeron. Predictions for HERA}
KFA-IKO(Th)-1994-37
DFTT 42/94, October 1994.

\smallskip

\item {[10]} M. Froissart,  Phys. Rev.  {\bf 123} (1961) 1053. For the proof of
this theorem starting from axiomatic field theory, see A. Martin,
\NC{42}{1966}{930}. The unexperienced reader who would like a more exhaustive
picture on this and related subjects, may profitably consult  R. J. Eden,
{\it High Energy Collisions of Elementary Particles} Cambridge Press (1967).

\item{[11]} a)  P. Desgrolard {\it et al.} \PL{B309}{1993}{191};

\item {}b) M. Bertini, M. Giffon and L. Jenkovszky,
{\it Small-$x$ behaviour of the Proton Structure Function}
to be published in the proceedings of VI$^{th}$ Rencontre de Blois,
{\it The heart of the matter}, France (June 1994), Edition Frontihres.
\smallskip

\item {[12]} A. Donnachie and P. V. Landshoff, Nuc. Phys.
{\bf B 244} (1984)
322 and Nuc. Phys. {\bf B 267} (1986) 690.

\item {[13]}
A.D. Martin, W.J. Stirling and  R.G. Roberts, {\it Parton Distributions of the
Proton}

RAL-94-055 DTP/94/34, June 1994 (and references therein).
\smallskip

\item {[14]}E.A. Kuraev, L.N. Lipatov and V.S Fadin,  \JETP{45}{1977}{199};

\item{} Ya. Ya. Balitsky and L.N. Lipatov,  \SJNP{28}{1978}{822};

\item{} J. C. Collins and  J. Kwiecinski,  \NP{B316}{1989}{307};

\item{} J. Kwieci\'nsky, A.D. Martin and  W.J. Stirling, \PR{D42}{1990}{3645};

\item{} J.Bartels, \NPps{B}{12}{1990}{201};

\item{} J. C. Collins and P. V. Landshoff, Phys. Lett. {\bf B 276} (1992) 196;

\item{}A.H Mueller,  J. Phys G {\bf 19} {\it Workshop on HERA - the New
Frontier
for QCD} (1994) 1463.
\smallskip

\item {[15]} R.D Ball and S. Forte,  \PL{B335}{1994}{77};

\item{} R.D Ball and S. Forte,  {\it The Rise of ${F}_2^p$ at HERA} to be
published in the proceedings of VI$^{th}$
Rencontre de Blois {\it The heart of the matter}, France (June
1994), Edition Frontihres.
\smallskip

\item {[16]} A. Capella, A. Kaidalov, C. Merino, and J. Tran Thanh Van,
Phys. Lett. {\bf B337}, (1994), 358.
For related approaches to this problem, the reader could profitably
refer to :
H. Abramowicz, E.M. Levin, A. Levy and U. Maor \PL{B269}{1991}{465};
A. Levy {\it The energy behaviour of the  real  and virtual photon-proton
cross sections}, DESY Report 95-003;
C. Bourrely, J. Soffer and T.T. Wu \PL{B339}{1994}{322}.
\smallskip

\item {[17]} NMC Collaboration, P. Amaudruz {\it et al.} \PL{B295}{1992}{159}.

\vfill\eject


\vglue 0.5 truecm

\centerline {\bf Figures captions}
\bigskip

\noindent {\bf {Fig. 1}} Kinematic and variables of the process
$l+N \rightarrow l^{'}+N^{'}+X$ used in the text.

\bigskip

\noindent {\bf Fig. 2} The fit of Eq. (7) ( obtained from Ref. 11a)
to the early HERA data extrapolated to the very small $x$ values.

\bigskip

\noindent {\bf Fig. 3} Small-$x$ structure function ${F}_2^p$ from H1 data [1]
(triangulated dots)
and ZEUS data [2] (closed points and stars) plotted as function of $x$
at fixed $Q^2$ compared
with the fit of Eq. (9) (solid line). Only data with $x \le 5.10^{-3}$ have
been used in the fit. The NMC data [17] (open points) are not fitted.

\bigskip

\noindent {\bf Fig. 4 a,b} Structure function with the same data of
{\bf Fig.3} plotted as a function of
$x$ at $Q^2$ fixed ({\bf a}) and as a function of $Q^2$ at $x$ fixed ({\bf b)}.
The solid line is obtained with Eq. (12). Only data with $x \le 10^{-2}$ have
been used in the fit.

\end